\documentclass[a4paper]{spie}  

\usepackage{algpseudocode}
\usepackage{algorithm}
\usepackage[utf8]{inputenc}
\usepackage{amsmath,amsfonts,amssymb}
\usepackage{graphicx}
\usepackage{xcolor}
\usepackage{soul}
\usepackage{pgfplots} 
\pgfplotsset{compat=newest} 
\pgfplotsset{plot coordinates/math parser=false} 
\newlength\figureheight 
\newlength\figurewidth 
\usepackage[colorlinks=true, allcolors=blue]{hyperref}

\title{Localization of Sound Sources in a Room with One Microphone}

\author{Helena Peić Tukuljac, Hervé Lissek and Pierre Vandergheynst}
\affil{{Signal Processing Laboratory LTS2}\\
École Polytechnique Fédérale de Lausanne (EPFL), CH-1015 Lausanne, Switzerland}

\authorinfo{Further author information: \\Helena Peić Tukuljac: helena.peictukuljac@epfl.ch \\  Hervé Lissek: herve.lissek@epfl.ch \\ Pierre Vandergheynst: pierre.vandergheynst@epfl.ch}

\begin{document} 
\maketitle

\begin{abstract}
Estimation of the location of sound sources is usually done using microphone arrays. Such settings provide an environment where we know the difference between the received signals among different microphones in the terms of phase or attenuation, which enables localization of the sound sources. In our solution we exploit the properties of the room transfer function in order to localize a sound source inside a room with only one microphone. The shape of the room and the position of the microphone are assumed to be known. The design guidelines and limitations of the sensing matrix are given. Implementation is based on the sparsity in the terms of voxels in a room that are occupied by a source. What is especially interesting about our solution is that we provide localization of the sound sources not only in the horizontal plane, but in the terms of the 3D coordinates inside the room.
\end{abstract}


\section{Introduction} \label{Introduction}
\label{sec:intro}  

In the last decade the theory of compressed sensing \cite{DonohoCS, CandesCS} has arised in the domain of acoustic signal processing. There was always a need for finding a structure in the high dimensional acoustical data that was cumbersome to handle. In 2015 Boche et al. \cite{CSAPP} provided a detailed state of the art for the application of compressed sensing in the domains of image and acoustic signal processing.
\par The origins of sparsity in acoustical data include, but are not limited to: voxels (directions of arrival) occupied by sound sources which is usually exploited for the localization of the sound sources in free field \cite{CompressiveBeamforming} by designing a Fourier domain dictionary, or in rooms for the estimation of the sound pressure distribution \cite{HearingBehindWalls}. The sparsity in the image-source model has been mainly used for the estimation of the room shape \cite{RoomShape} and the estimation of the direction of arrivals of early echoes \cite{EchoLocalization}, the sparsity of plane wave representation of the sound pressure is used for the characterization of the sound fields inside the room \cite{WaveRepresentation}. The sparsity of the room modes may also be exploited in the low-frequency range of the room transfer functions (RTF) \cite{Low}. By RTF we denote the ration between the received and emitted signal in Fourier domain.

\par Instead of estimating the position of the sound sources from time difference of arrival between different microphones in an array \cite{VehicleLocalization, BimodalLocalization}, we aim to rely only on one microphone and combine the sparsity that exists in the term of the voxels of a room occupied by the sound sources and the low-frequency room modes in the RTF toward successfull localization. To this end, we will analyze the transfer functions below the so called Schröeder frequency, which is defined as: $f_s = 2000 \sqrt{\frac{RT_{60}}{V}}$, where $V$ is the volume of the room and $RT_{60}$ is the reverberation time \cite{RoomAcoustics}. This combination should result in a fast localization of sound sources by only one microphone as will be further explained.

The remainder of the paper is organized as follows: In Section \ref{RIR} we discuss the sparsity that exists in the low frequency domain of room transfer functions. Section \ref{CompressedSensing} gives a general introduction to compressed sensing and its application to the localization of sound sources. The design and the limitations of the sensing matrix for our case is given in Section \ref{SensingMatrix} and final remarks and conclusions are given in Section \ref{Conclusion}.

\section{Modal representation of the sound pressure and its Low-frequency Properties} \label{RIR}
In the further development of our approach, we are going to rely on two facts: the room shape is known and the microphone position is known. These assumptions imply that we know the resonant frequencies of the room and the room modes related to the microphone's positions.
\par The solution of the wave equation for the sound pressure at a given receiver position $\mathbf{r_{mic}}$ and for a given source position $\mathbf{r_{ss}}$ in a room in the Fourier domain (with the angular frequency $\omega$) is given by \cite{RoomAcoustics}:
\begin{equation} \label{eq:1}
H_{\omega}(\mathbf{r}_{\textrm{mic}},\mathbf{r}_{\textrm{ss}})=\rho_0 c^2 \omega Q \sum_{n}\frac{\Xi_n(\mathbf{r}_{\textrm{mic}})\Xi_n(\mathbf{r}_{\textrm{ss}})}{K_n [2\delta_n \omega_n + i(\omega^2 - \omega_n^2)]}
\end{equation}
where $\rho_0$ is the density of the propagating medium (air), $c$ is the sound celerity, $Q$ is the volume flow velocity of the sound source, $\Xi_n(\cdot)$ are the eigenfunctions, $K_n$ is the gain, $\delta_n$ is the damping coefficient and $\omega_n$ are the resonant frequencies. We can notice an interesting underlying symmetry that exists in this equation: position of the microphone $\mathbf{r}_{\textrm{mic}}$ and position of the sound source $\mathbf{r}_{\textrm{ss}}$ are interchangeable, meaning that if we exchange these positions, the expression will remain the same.

\begin{figure}[h]
\includegraphics[width=\columnwidth]{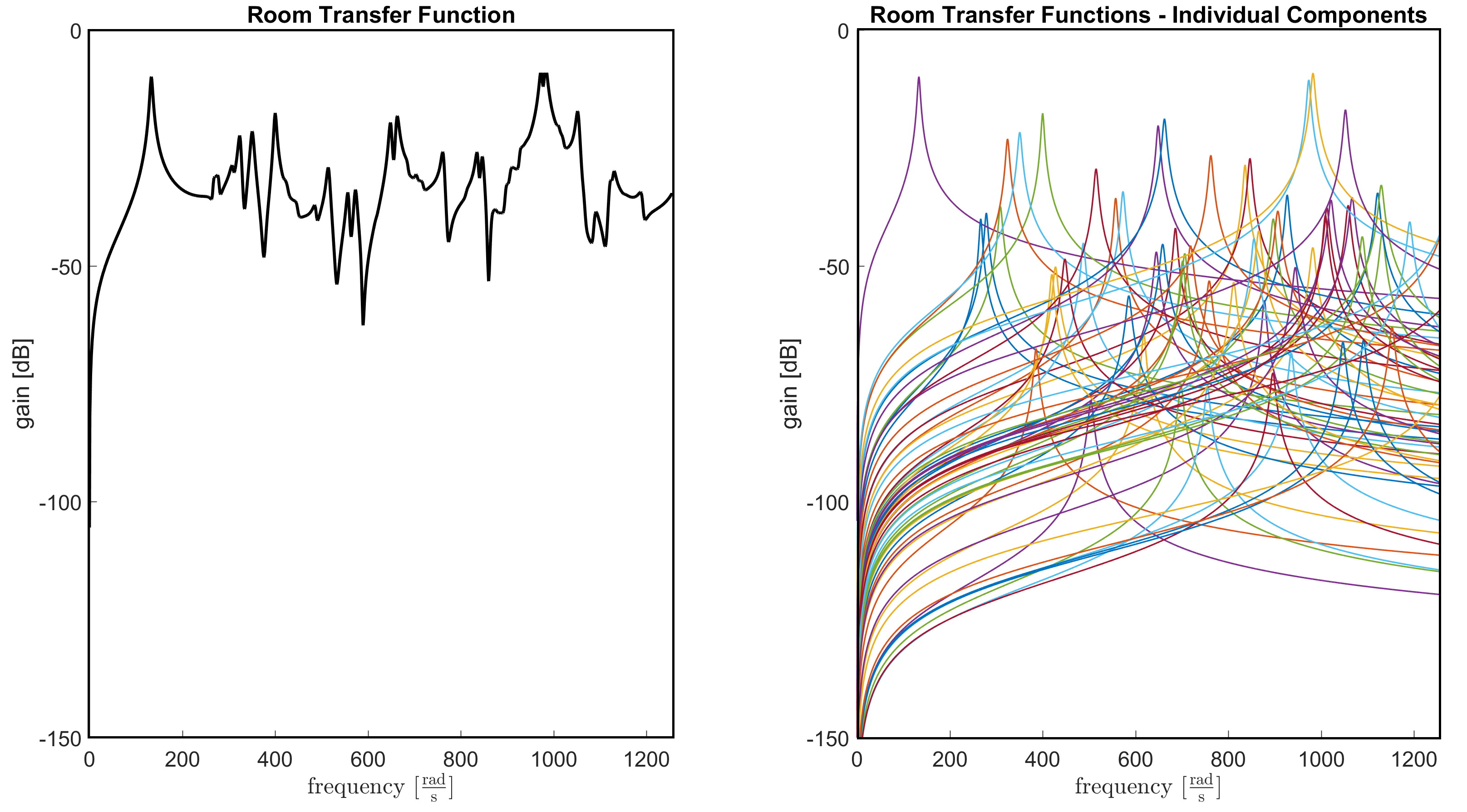}
\caption[width=\textwidth]
{ \label{fig:room_modes} Individual components of the RTF are called room modes. As illustrated, room modes can be simply modeled as second order bandpass filters.}
\end{figure}
In Figure \ref{fig:room_modes} we can see a segment of RTF for an arbitrary set of positions $\mathbf{r}_{mic}$ and $\mathbf{r}_{ss}$ up to $200Hz$ ($1200\frac{rad}{s}$) and its decomposition into the room modes. The sharpness of the peaks of room modes is dependent on the damping properties of walls of the room. Peaks of the room modes are aligned with the resonant frequencies of the room.

The angular eigenfrequencies for a rectangular room of size $L_x \times L_y \times L_z$ are given by the expression: $\omega_r = \pi c \sqrt{\big(\frac{n_x}{L_x}\big)^2+\big(\frac{n_y}{L_y}\big)^2+\big(\frac{n_z}{L_z}\big)^2}$ where $(n_x, n_y, n_z) \in \mathbb{N}_0^3 \setminus (0, 0, 0)$.

\subsection{Room Transfer Function at Different Positions Across the Room}
In 1985, Richardson et al. \cite{CurveFitting} have proposed a curve fitting algorithm allowing the reconstruction of the RTF curve from discrete measurements using room mode shaped functions as basic fitting elements. Each RTF is characterized by a set of parameters: resonant frequencies (eigenfrequencies) which are aligned with the position of the peaks of room modes, and with damping, attenuation and phase of these room modes.

For different positions of the microphones/sound sources across the room, some parameters stay the same - \textit{common parameters}: eigenfrequencies which depend on the room shape, and the room mode damping which depends on the damping of the wall. The attenuation and the phase of the room modes are position dependent parameters - \textit{specific parameters}.

\begin{figure}[H]
\includegraphics[width=\columnwidth,height=11cm]{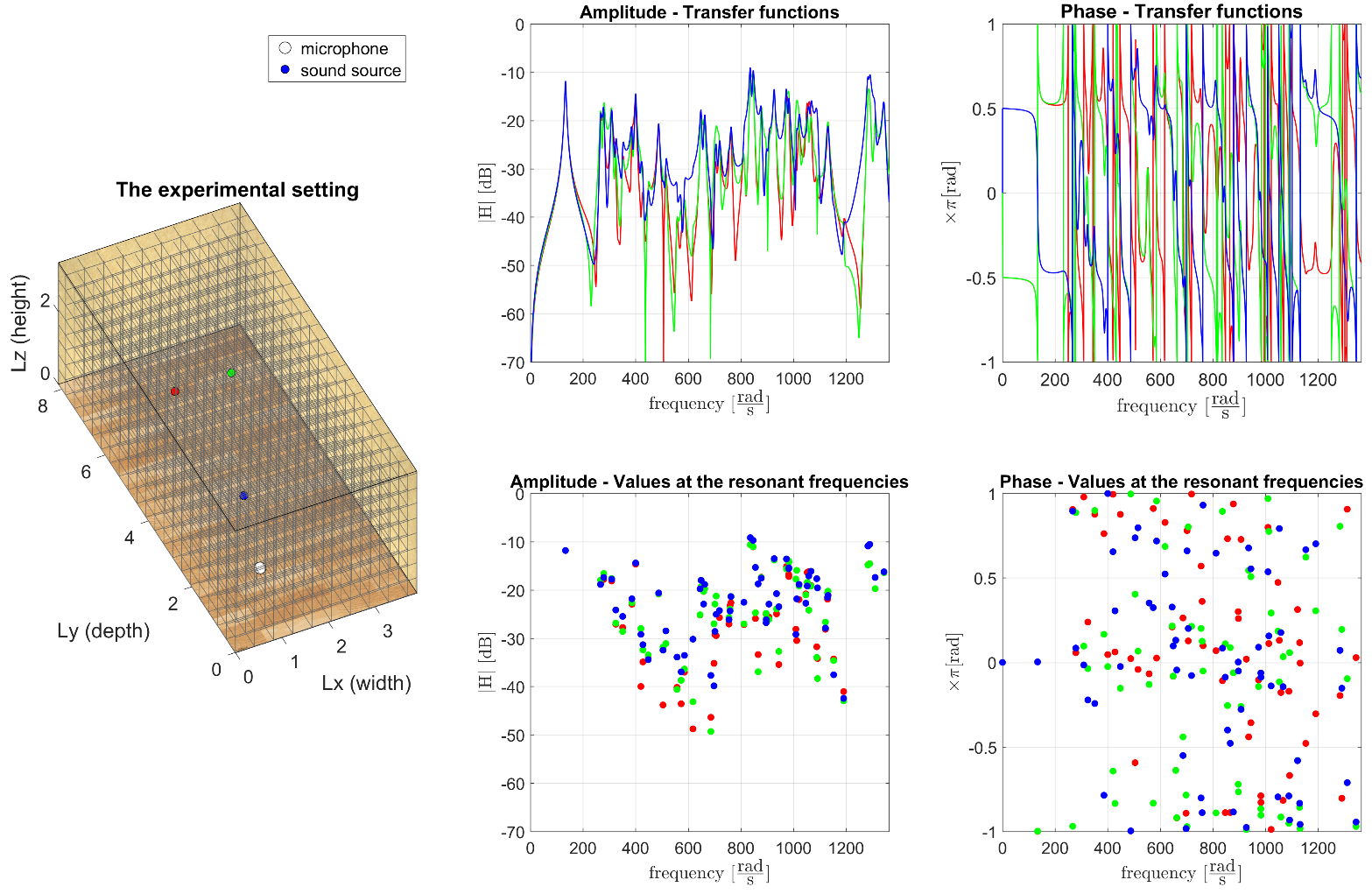}
\caption[width=\textwidth]
{\label{fig:transfer_functions} Values of the RTF across the room vary in the terms of attenuation and phase value at the resonant frequencies. We exploit only the difference in the attenuation because in our target experimental setting there exists only one microphone and the sources will emit white noise.}
\end{figure}

Figure \ref{fig:transfer_functions} illustrates the difference between the attenuation and the phase of the RTFs across the room at the resonant frequencies. White point shows the fixed and known position of the microphone and colorful points are the positions of sound sources that should be estimated. As can be observed, although that all the positions of the sound sources result in the peaks at the same set of frequencies (the resonant frequencies of the room), the set of the heights of these peaks seems unique (this will be further observed in the next section). This means that each pair of the positions of a sound source and a microphone could potentially result in a unique set of attenuation factors at the resonant frequencies.

Although that there exists uniqueness of phase for each room mode, since we plan to use only one microphone and white noise sources, this is irrelevant for our case but has a potential for some other type of room characterization. We have decided to investigate the potential of unique representation of position of the sound source within the room with the set of attenuations of RTF at resonant frequencies. Therefore we have established a valuable reasoning for the design of our sensing matrix.

\subsection{Relation Between Room Modes and Plane Waves}
In a rectangular room, each eigenfunction (eigenmode of the Laplacian operator) represents a sum of 8 plane waves that share a wave number:
\begin{equation}
\Xi(\mathbf{k}_n,\mathbf{r}_{\textrm{m}})= \sum_{i=1}^8 a_i e^{j(\mathbf{S}(:,j) \odot \mathbf{k}_n) \cdot \mathbf{r}_m}
\end{equation}
where $\odot$ is a Hadamard product, $\mathbf{S}_{3 \times 8}$ is a sign matrix whose columns alternate from $[1,1,1]^T$ to $[-1,-1,-1]^T$, $\mathbf{k}_n=(\frac{n_x \pi}{L_x}, \frac{n_y \pi}{L_y}, \frac{n_z \pi}{L_z})$, $(n_x, n_y, n_z) \in \mathbb{N}_0^3 \setminus (0, 0, 0)$, is the eigenvalue of the wave equation for the $n^{\mathrm{th}}$ room mode (wave vector), and $\mathbf{r}_m$ is a position inside the room.

\begin{figure}[H]
\includegraphics[width=\textwidth,height=4cm]{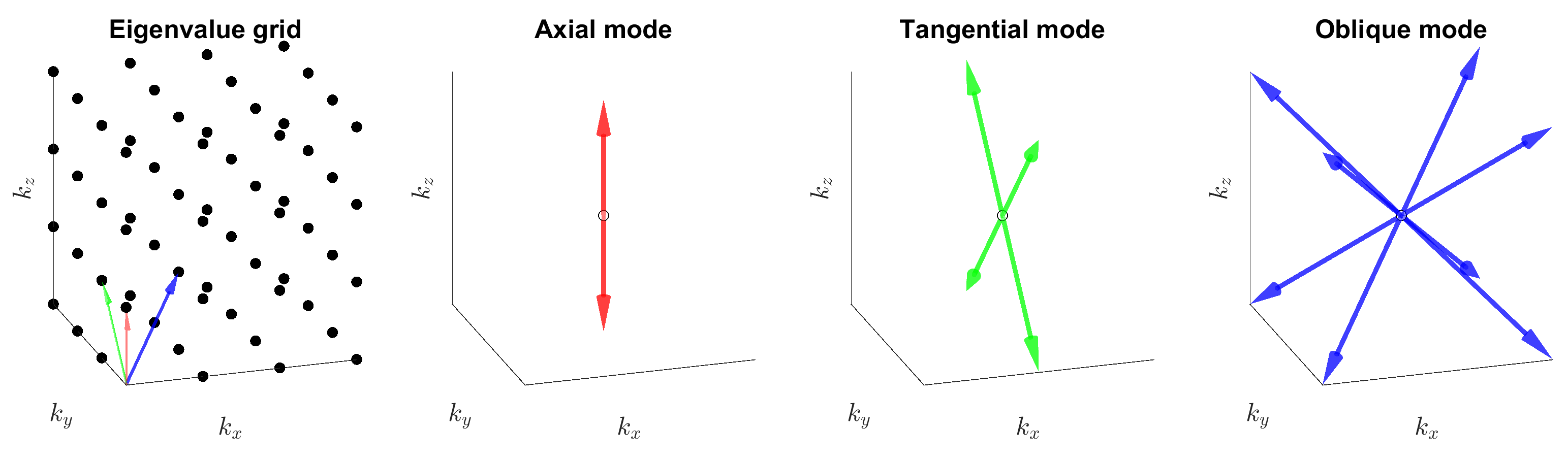}
\caption[width=\textwidth]
{ \label{fig:mode_grid} Eigenvalue space of a rectangular room with rigid walls. The left-hand side shows just one octant because of the symmetry that exists (there are 8 plane waves for each wave number). The length of the wave vector is proportional to the eigenvalue of the Laplacian.}
\end{figure}

As can be seen in Figure \ref{fig:mode_grid}, these wave vectors are just corners of a parallelepiped ($\mathbf{k}=[\pm k_x,\pm k_y, \pm k_z]^T$). 
We can also notice the periodicity of the wave vector grid: $\frac{\pi}{L_x}$, $\frac{\pi}{L_y}$, $\frac{\pi}{L_z}$, along each of the axes.

\begin{figure}[H]
\includegraphics[width=\textwidth,height=8cm]{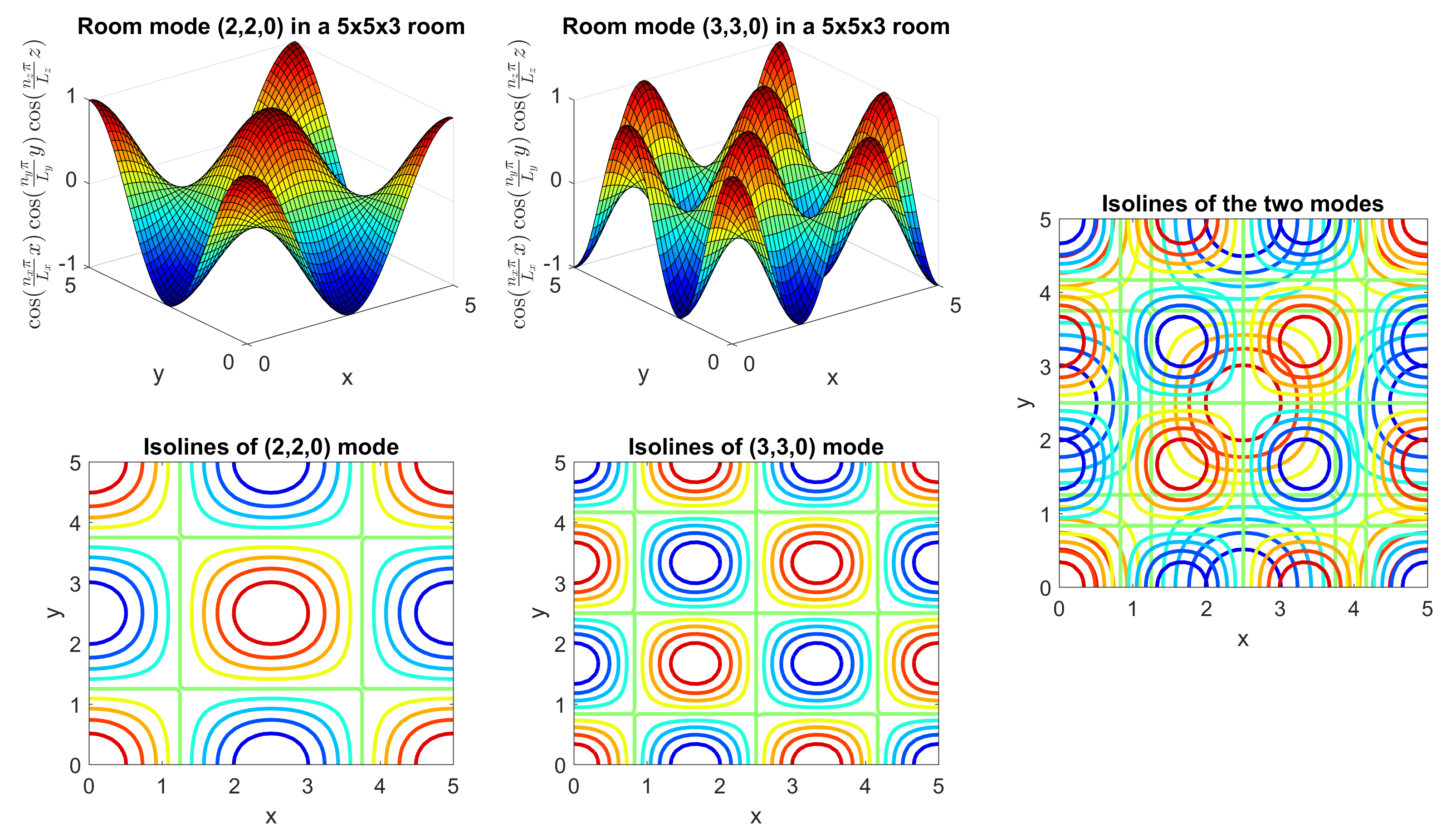}
\caption[width=\textwidth]
{\label{fig:room_mode_rigid_wall}An example of $(n_x, n_y, n_z) \in \{(2,2,0), (3,3,0)\}$ room modes in a $5\mathrm{m} \times 5\mathrm{m} \times 3\mathrm{m}$ room with rigid walls. We can notice that the isolines of different modes intersect in just a few locations, which supports our assumption of different height of sets of peaks in the RTF.}
\end{figure}

In a theoretical case for which all walls of a room are perfectly rigid, all plane waves have the same expansion coefficient ($\forall i,\;  a_i = a$), so our sum of the 8 plane waves can be represented as a product of cosine functions:
\begin{equation}
\Xi(\mathbf{k}_n,\mathbf{r}_m) \sim a \cos{\Big(\frac{n_x \pi}{L_x}\mathbf{r}_m(x)\Big)}\cos{\Big(\frac{n_y \pi}{L_y}\mathbf{r}_m(y)\Big)}\cos{\Big(\frac{n_z \pi}{L_z}\mathbf{r}_m(z)\Big)}.
\end{equation}
where $\mathbf{r}_m(x),\;\mathbf{r}_m(y),\;\mathbf{r}_m(z)$ are the Cartesian coordinates of position $\mathbf{r}_m$ and $a$ is a constant. An example of room mode for a room with rigid walls is given in Figure \ref{fig:room_mode_rigid_wall}.

\subsection{Ambiguities that Exist in the Terms of Uniqueness of the Attenuation Across the Room}

\begin{figure}[H]
\begin{center}
\includegraphics[width=\textwidth]{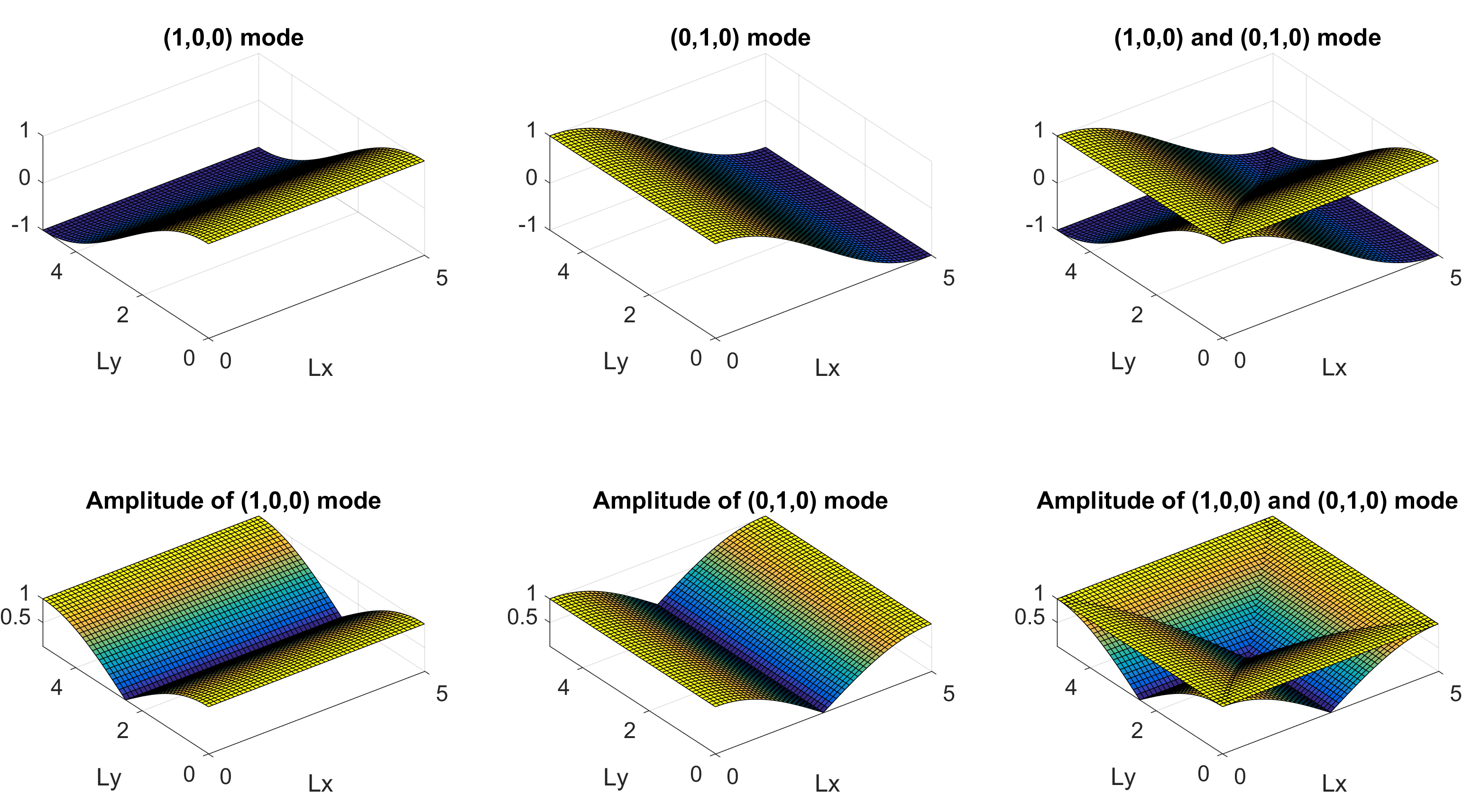}
\end{center}
\caption[width=\textwidth]
{\label{fig:basic_modes} Basic modes and their attenuation values.}
\end{figure}

We will observe the basic axial modes in Figure \ref{fig:basic_modes} in order to illustrate that relying only on them would not be sufficient to have a unique position representation. Although that the sound pressure value function is in 5D, we can successfully visualize only values in 3D. First row shows the $x$- and $y$-axial modes (everything that will be said applies analogously to $z$-axial modes as well). We can see that these two modes form pairs of points that result in a unique location identifier. But, since we have decided to explore the special case with only one microphone, we need to neglect the phase of the RTF, therefore we can just observe the absolute value of the RTF. As seen in the second row of the same figure, this introduces ambiguity - there exists a unique representation, but only in $\frac{1}{8}$ of the room. 

\begin{figure}[h]
\begin{center}
\includegraphics[width=11cm,height=6.5cm]{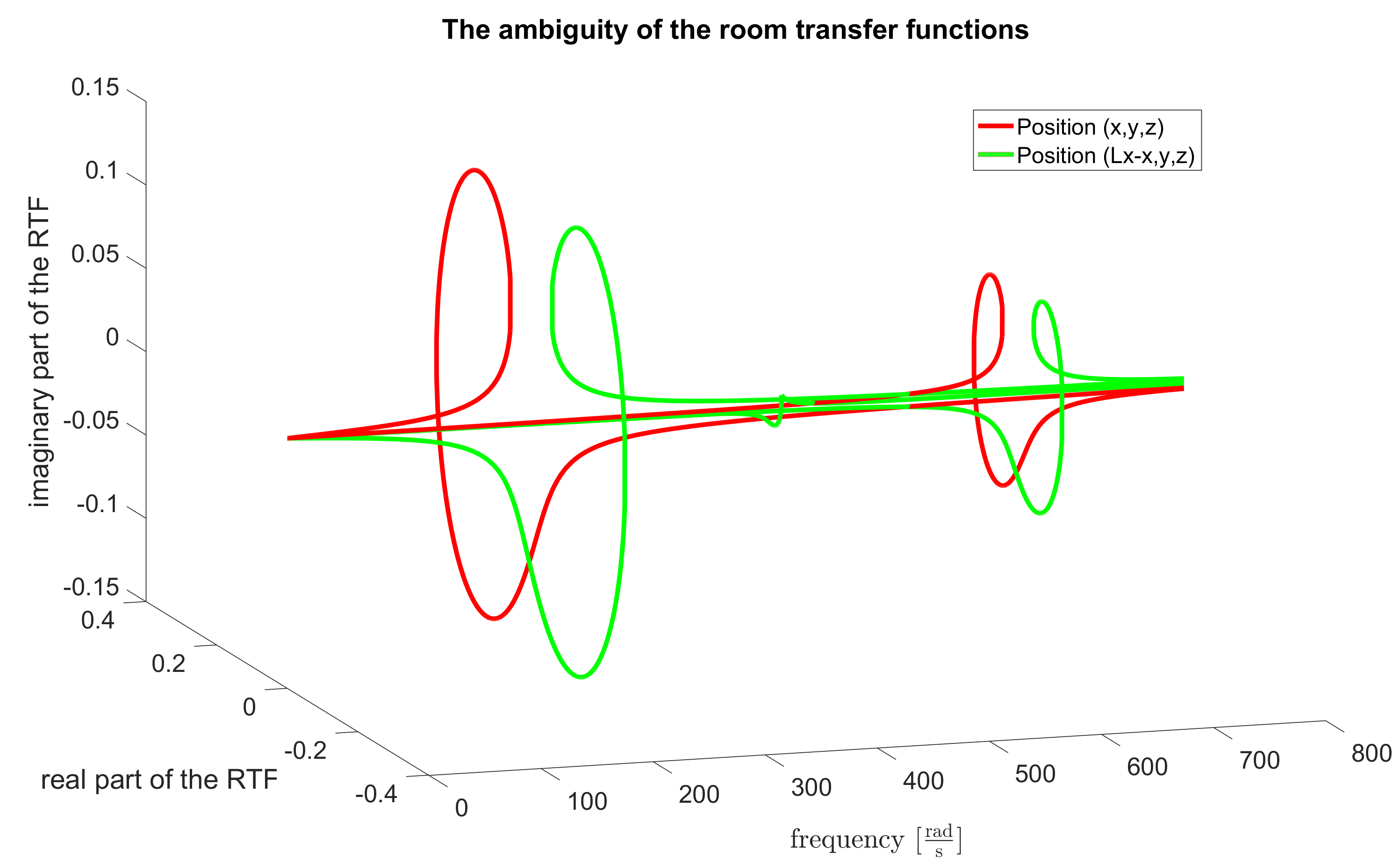}
\end{center}
\caption[width=\textwidth]
{\label{fig:ambiguity_in_room_transfer_function} Ambiguities that exist in the term of the uniqueness of the RTF.}
\end{figure}

This ambiguity is illustrated in Figure \ref{fig:ambiguity_in_room_transfer_function}. Axis represent the dependency between frequency and the real and the imaginary part of the RTF. Here we see 3 modes of two different positions in a room. The small mode in the middle has the same amplitude and phase and the other two modes have an opposite phase, which can not be seen when we project it to neglect the phase. This means that we can not rely only on the basic axial modes for the sound source localization.

\section{Compressed Sensing and Sound Source Localization} \label{CompressedSensing}
\subsection{Sparse Representation of the Position of Sources}
In sound source localization problems the domain of interest is usually divided into an angular grid such that the sources occupy just a few of these angles. Since our sources are positioned inside a room, we will divide the room into voxels and assume that the number of voxels occupied by a source is small. We recognize that this is a problem with underlying sparsity. These problems are usually solved by using the theory of compressed sensing.

\subsection{Compressed Sensing}
Our signal of interest $\mathbf{y}$ is the measurement of sound pressure at a known location inside a known room:
\begin{equation}
\mathbf{y=\Psi \Phi x}
\end{equation}
where $\mathbf{y} \in \mathbb{R}^N$ are the sound pressure measurements, $\mathbf{\Psi}_{N \times N}$ is the inverse Fourier Transform (represents the change of domain), $\mathbf{\Phi}_{N \times M}$ is a representational dictionary (columns of this matrix are called atoms) with the RTFs as columns and $\mathbf{x} \in \mathbb{R}^M$ are the sparse expansion coefficients. The product $\mathbf{A} = \mathbf{\Psi}\mathbf{\Phi}$ is usually referred to as the sensing matrix. $\mathbf{x}$ is $K$-sparse, which means that it contains at most $K$ non-zero elements and $K \ll N$. Since $M > N$ we are facing an underdetermined system of equations. The problem in this form is non-convex. Introducing the assumption on the sparsity of $\mathbf{x}$ makes the problem well-posed.

These types of problems are usually solved using one of the 5 groups of approaches listed in \cite{SparseCompare}, where the most common ones are the convex relaxation and the greedy pursuit. The convex relaxation \cite{ConvexRelaxation,Boyd}, which is also known as the Basis Pursuit, relies on the relaxation of the minimization of the $\ell_0$-norm to $\ell_1$-norm which favors the sparse solutions, although at a cost of requiring higher number of measurements \cite{l1morel0}. There also exists a relaxation to $\ell_2$-norm, but this norm favors the minimization of the energy of the signal rather than finding a sparse solution. In practice convex relaxation approach is usually used for smaller and medium size problems, because large scale data causes computational issues.

In our solution we will rely on the greedy approaches such as Orthogonal Matching Pursuit (OMP) \cite{OMP} and Compressive Sampling Matching Pursuit (CoSaMP) \cite{CoSaMP}. These methods select up to K atoms of a dictionary that give the least approximation error. CoSaMP is a faster contemporary method which works by selecting multiple atoms at every iteration. The main drawback of these methods is that the sparsity of the signal has to be known upfront.

Regardless of the approach, one of the main advantages of compressed sensing technique is the robustness to noise since we project our signal to the vectors that span the signal space, and therefore we neglect the residual related to the existing noise as long as the noise is not highly correlated with the signal.

\subsection{Conditions for Dictionary Design}
\subsubsection{Spark and coherence of the dictionary}
Spark of a matrix $\mathbf{\Phi}$ is the smallest number of linearly dependent columns of matrix $\mathbf{\Phi}$. The requirement for the sensing matrix $\mathbf{\Phi}$ in compressed sensing is that the following holds:
\begin{equation}
\mathrm{spark}(\mathbf{\Phi}) > 2K
\end{equation}
where $K$ is the level of sparsity. In other words: To achieve an injective mapping we need to assure that there are no two $K$-sparse vectors that map to the same measurements. This implies that the rank of our sensing matrix has to be at least $2K$ which is tightly related to the restriction on the coherence of the dictionary. 

According to the theory of compressed sensing, we have to ensure the appropriate coherence parameter of a dictionary: $\mu =\max_{1\leq i<j\leq n}\frac{|\langle \mathbf{\varphi}_i, \mathbf{\varphi}_j \rangle|}{\lVert \mathbf{\varphi}_i\rVert_2\lVert\mathbf{\varphi}_j\rVert_2}=\max_{1\leq i<j\leq n}\cos{\sphericalangle(\mathbf{\varphi}_i,\mathbf{\varphi}_j)}$. Coherence is the cosine of the acute angle between the closest pair of atoms in a given dictionary. We want our dictionary to be incoherent, so $\mu$ should be the smallest possible. The best case is the case where we have orthogonal atoms with the coherence parameter equal to zero between different atoms of the dictionary.

\subsubsection{Restricted Isometry Property}
The restricted isometry property guarantees that the distances (lengths) are preserved when moving from one space to another.
Let $\Phi$ be an $M \times N$ matrix and let $1 \leq K \leq N$ be an integer. Suppose that there exists a constant $\delta_{K} \in (0,1)$ such that, for every $M \times K$ submatrix $\mathbf{\Phi}_K$ of $\mathbf{\Phi}$ and for every $K$-sparse vector $\mathbf{y}$,
\begin{equation}
(1-\delta _{K})\|\mathbf{y}\|_{2}^{2}\leq \|\mathbf{\Phi}_K \mathbf{y}\|_{2}^{2}\leq (1+\delta _{K})\|\mathbf{y}\|_{2}^{2}.
\end{equation}
Then, the matrix $\mathbf{\Phi}$ is said to satisfy the $K$-restricted isometry property with restricted isometry constant $ \delta_{K}$. In most cases it is hard to check whether this property holds or not.

\subsection{Sound Source Localization in a Room}
The following question rises: How to tailor a simple incoherent dictionary for fast localization of sources inside the room? In order to have a well-posed problem we introduce the following assumptions:
\begin{enumerate}
\item the shape of the room and the reverberation time are known,
\item the position of the microphone is known, and
\item all the sound sources have a flat spectrum in the observed frequency range.
\end{enumerate}

In most cases the Restricted Isometry Property is hard to check. We know that random matrices, which were used as the dictionaries in the early stages of compressed sensing, satisfy this property. Therefore we will choose the potential position of the sources on uniformly at random on the regular grid.

For each of the potential positions of sound sources and a fixed position of the microphone we have one atom in the dictionary which consists out of the height of the peaks in the RTFs at the resonant frequencies. The height of the dictionary is proportional to the number of the resonant frequencies in the observed frequency range. The number of resonant frequencies below a given frequency $f_s$ \cite{RoomAcoustics} can be computed by: $N(f_s)=\frac{4}{3}\pi V \big(\frac{f_s}{c}\big)^3+\frac{1}{4}\pi S \big(\frac{f_s}{c}\big)^2+\frac{1}{2}L\frac{f_s}{c}$, where $V = L_xL_yL_z$, $S=2(L_xL_y+L_yL_z+L_zL_x)$ and $L=L_x+L_y+L_z$. The width of the dictionary is proportional to the number of observation points on the predefined grid.

In order to localize the sources, we search for a subset of atoms that give the best fitting for the signal recorded by the microphone. Once we discover which atoms of our sensing matrix have the highest expansion coefficients in the sparse representation, we can easily recover the position of the sound sources in the room, because we know which atom corresponds to which position since we have tailored the dictionary ourselves.

\section{Designing an Efficient Sensing Matrix}
\label{SensingMatrix}
\subsection{Coherence}
Coherence of a dictionary can be seen from the maximum off-diagonal element of the coherence Gram matrix $\mu = \max_{i \ne j} \mathbf{G}_{ij}$. In our case where $\mathbf{\Psi}_{N \times N}$ is the inverse Fourier Transform and $\mathbf{\Phi}_{N \times M}$ is the matrix with the RTF coefficients, the Gram matrix has the following form:
\begin{equation}
\mathbf{G}=|\mathbf{A}^H \mathbf{A}|=|\big(\mathbf{\Psi}\mathbf{\Phi}\big) ^H\mathbf{\Psi}\mathbf{\Phi}|=|\mathbf{\Phi}^H\mathbf{\Psi}^H\mathbf{\Psi}\mathbf{\Phi}|.
\end{equation}
Since the Fourier matrix has orthonormal atoms up to a scaling constant $\mathbf{\Psi}^H\mathbf{\Psi}=\frac{1}{N}\mathbf{I}$, we have:
\begin{equation}
\mathbf{G}=\frac{1}{N}\mathbf{\Phi}^H\mathbf{\Phi}.
\end{equation}

Therefore we observe the coherence of the sensing matrix by focusing on the discretization of the room transfer function.
Since our exponentials in the plane wave representation are not equidistant, we can not apply the Dirichlet kernel sum to our case to simplify the expression (an approach common for many solutions \cite{FRI, CompressiveBeamforming, BasisMismatch}).

For a uniform case, the off-diagonal elements of our Gram matrix are proportional to:
\begin{equation}
\mathbf{G}_{ij} \sim 
    cos(k_x r_x)cos(k_x (r_x\pm m \Delta x))+
    cos(k_y r_y)cos(k_x (r_y\pm n \Delta y))+
    cos(k_z r_z)cos(k_x (r_z\pm o \Delta z)).
\end{equation}
It results in a complex form of the elements of Gram matrix. Some observations have shown that we are dealing with highly correlated atoms. Therefore we need to find a workaround in order to have a successful source localization. 
Due to the smoothness of cosine function, the points on the potential sound source position grid that lay close, result in similar heights of the peaks in RIR. 

\subsection{Battle of the Grids}
\begin{figure}[h]
\includegraphics[width=\textwidth]{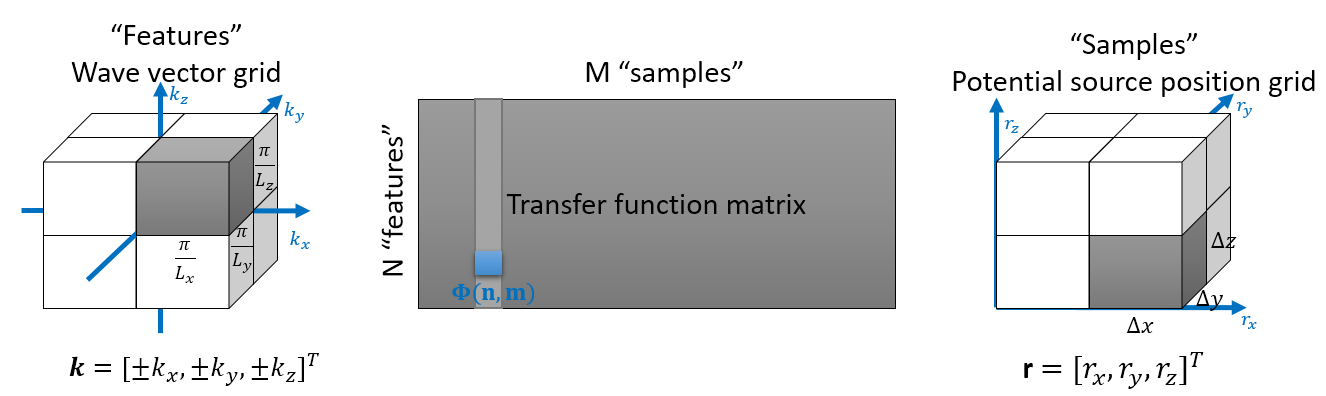}
\caption[width=\textwidth]
{ \label{fig:battle_of_the_grids} Two grids that represent two degrees of freedom that we have for designing the sensing matrix.}
\end{figure}

Our problem has two degrees of freedom and both of them represent a selection process of the nodes on a uniform grid. We have a grid of wave vectors - \textit{features} and a grid of potential positions of sound sources - \textit{samples}.
In Figure \ref{fig:battle_of_the_grids} the grid on the left-hand side repeats in all 6 directions and the one on the right-hand side repeats in 3 directions.

We will observe the room transfer function in a matrix form at the resonant frequencies. If we go back to equation (\ref{eq:1}) and introduce $\omega=\omega_n$, we get that each of the elements of our sensing matrix $\mathbf{\Phi}_{N \times M}$ is of the following form:
\begin{equation}
\color{cyan}\mathbf{\Phi}(n,m)\color{black}= \frac{\rho_0 c^2 \color{cyan}Q_m\color{black}}{2 K_n \delta_n}\Xi(\mathbf{k}_n,\mathbf{r}_{\textrm{mic}})\color{cyan}\Xi(\mathbf{k}_n,\mathbf{r}_{m})\color{black}
\end{equation}
which corresponds to $n^{\textrm{th}}$ wave vector and $m^{\textrm{th}}$ potential sound source position. The only coefficients that differ among the atoms of the dictionary are represented in blue. The difference due to the volume velocity of the sound source $Q$, will not affect our approach since we assume that we are observing our sound sources in a linear regime. This parameter has an effect only on the expansion coefficients of the sparse representation. Therefore we focus on the sound sources' position that produces different attenuation of room modes.

So the RTF matrix has the following decomposition:
\begin{equation}
\mathbf{\Phi}=\frac{\rho_0 c^2}{2}
\begin{bmatrix}
    \frac{\Xi(\mathbf{k}_1,\mathbf{r}_{\textrm{mic}})}{K_1 \delta_1} & \hdots  & \frac{\Xi(\mathbf{k}_1,\mathbf{r}_{\textrm{mic}})}{K_1 \delta_1} \\
       \vdots  &   \ddots  &   \vdots \\
    \frac{\Xi(\mathbf{k}_N,\mathbf{r}_{\textrm{mic}})}{K_N \delta_N}  & \hdots  & \frac{\Xi(\mathbf{k}_N,\mathbf{r}_{\textrm{mic}})}{K_N \delta_N}
\end{bmatrix}
\odot
\color{cyan}
\begin{bmatrix}
    Q_1 \Xi(\mathbf{k}_1,\mathbf{r}_{\textrm{1}})  & \hdots  & Q_M \Xi(\mathbf{k}_1,\mathbf{r}_{\textrm{M}}) \\    
    \vdots &   \ddots  &   \vdots \\
    Q_1 \Xi(\mathbf{k}_N,\mathbf{r}_{\textrm{1}})  & \hdots  & Q_M \Xi(\mathbf{k}_N,\mathbf{r}_{\textrm{M}})
\end{bmatrix}\color{black}.
\end{equation}

Just to recall, our rigid wall room modes are of the form: $\Xi(\mathbf{k}_n,\mathbf{r}_{\textrm{m}})= \sum_{i=1}^8 e^{j(\mathbf{S}(:,j) \odot \color{cyan}\mathbf{k}_n\color{black}) \cdot \color{cyan}\mathbf{r}_m\color{black}}$, where $\mathbf{k}_n$ belongs to the positive octant of the left-hand side grid.

\section{Results}

\subsection{The Recovery of Signal's Support in a Highly Coherent Dictionary}
Candès et al. \cite{CoherentDictionaryRecovery} discuss the potential of recovery of data that has a sparse representation in a coherent dictionary. Coherent dictionaries can give guarantees only on the recovery of the sparse signal, but not on the recovery of the set of indices of atoms in sparse representation. That is because if we have pairs of atoms that are extremely coherent (almost collinear), e.i. we are far away from satisfying $\mu \leq \frac{1}{3(s-1)}$, we can not tell which one of them will be used for our sparse representation when projecting to a lower-dimension space. Schnass et al. have approached this problem by introducing a complementary dictionary of the same size, but with low coherence, which maintains the sparse support of the measurements \cite{RedundantDictionaries}. Our approach will be in the spirit of random subdictionary selection \cite{RandomSubdictionaries}. There have been some approaches with subsampling of dictionaries over rows and columns in order to increase the speed of the convergence of greedy methods \cite{MPSto, StoCoSaMP}, but using such subsampling methods for coherent dictionaries is still unexplored. Authors of these papers named one of these methods as StoCoSaMP (Stochastic CoSaMP).

We restate our problem in the following manner: Recover sparse signal $\mathbf{x}$ from the following:
\begin{equation}
\mathbf{S}_{rf}\mathbf{\Psi}^*\mathbf{y} = \mathbf{S}_{rf} \mathbf{\Phi} \mathbf{S}_{sp} \mathbf{x}
\end{equation}
where $\mathbf{y}$ is the measured signal, $\mathbf{S}_{rf}$ is a resonant frequency selector that defines which points on the wave vector grid we observe, $\mathbf{S}_{sp}$ is a sound source position selector that defines which points on the potential source position grid we observe and $\mathbf{\Psi}^*$ is the Fourier transform. Both matrices,  $\mathbf{S}_{rf}$ and $\mathbf{S}_{sp}$, are just submatrices of an identity matrix - the first one is constructed from selected rows and the second one is constructed from selected columns. We could characterize our case as a highly sparse case, since the number of sources to be localized is going to be small (only one or a few).

Support of $\mathbf{x}$ shows which of the positions on the grid are the most probable positions of the sources. Without subsampling of the coherent dictionary, this support is usually wrongly estimated due to the ill-conditioness of the problem coming form the high coherence of the dictionary.

Here is the description of the algorithm ($\mathbf{I}$ is the identity matrix):

\algdef{SE}[DOWHILE]{Do}{doWhile}{\algorithmicdo}[1]{\algorithmicwhile\ #1}
\begin{algorithm}[H]
\begin{algorithmic}
\caption{Localization of sound sources in a room with one microphone}
    \State \textbf{Input}: Highly coherent room mode dictionary $\mathbf{\Phi}_{N \times M}$, uniform grid of potential points of the sound sources and ground truth positions of the sound sources (including the measured signal in Fourier domain $\mathbf{y^F} = \mathbf{\Psi^*}\mathbf{y}$).
    \State \textbf{Output}: Reconstructed positions of the sound sources.
    \Do
      \State Generate random subsampling matrices $\mathbf{S}_{rf} \underset{row}{\subset}\mathbf{I}_{N \times N}$ and $\mathbf{S}_{sp} \underset{column}{\subset} \mathbf{I}_{M \times M}$.
      \State Subsample the dictionary: $\Phi_{ss}= \mathbf{S}_{rf} \mathbf{\Phi} \mathbf{S}_{sp}$ and the measured signal $\mathbf{y^F_{ss}}=\mathbf{S}_{rf}\mathbf{y^F}$.
      \State Try to estimate the positions of the sound sources by estimating the support of $\mathbf{x}$ on $\mathbf{\Phi}_{ss}$ using CoSaMP for the given measured signal $\mathbf{y^F_{ss}}$ knowing the level of sparsity.
    \doWhile{CoSaMP \cite{CoSaMP} sparse representation does not converge (has norm of the residual a lot greater than zero)}
\end{algorithmic}
\end{algorithm}

\begin{figure}[H]
\begin{center}
\includegraphics[width=17cm, height = 6cm]{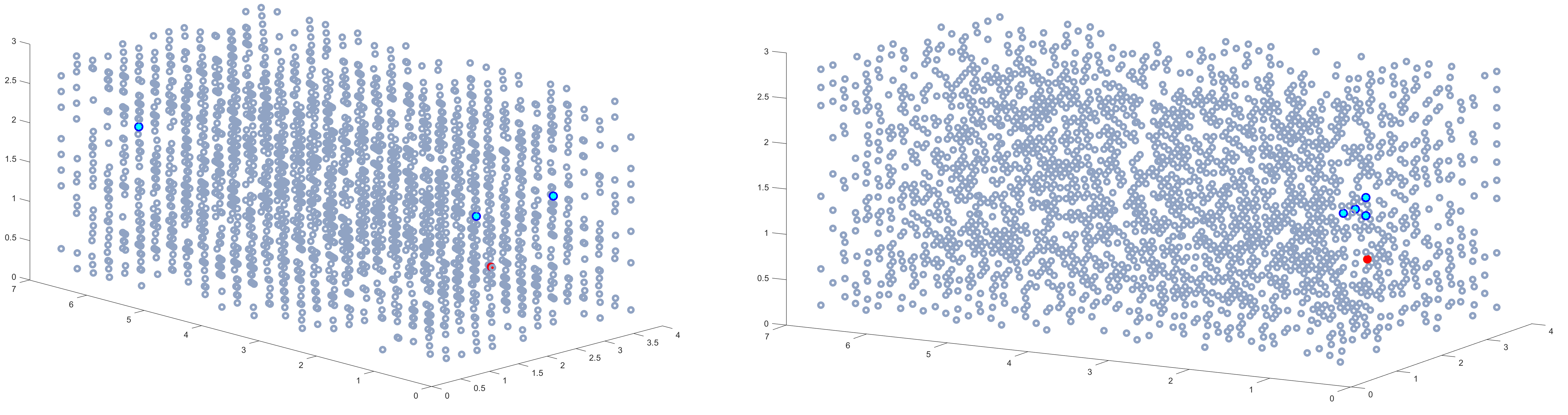}
\end{center}
\caption[width=\textwidth]
{ \label{fig:results} These are the results for localization of 3 sound sources inside a $4m \times 7m \times 3m$ room for a uniformly undersampled $10 \times 15 \times 10$ grid.}
\end{figure}

Figure \ref{fig:results}. shows a reconstruction example for a case with 3 sound sources. Grey circles are the potential positions taken into account in the current iteration, blue circles are the true positions and light blue points are the reconstructed positions. The red point represents the known position of the sound source. This algorithm has no problems with identifying position of sources that are close, as can be seen from the right hand side of the figure.

We will observe how different subsampling schemes effect the success and speed of our sparse support estimation.

We have performed 100 Monte Carlo simulations for each set of parameters and for the estimation of the position of two sound sources. Experiments were performed on a single core of Intel Xeon processor at 2.8GHz of a computer with 16GB of RAM. If the algorithm did not converge within 300 iterations, we would consider that to be a failure. If we do not bound the number of iterations, the algorithm always converges but sometimes it needs a few thousands of iterations. Reconstruction time does not include the time needed for constructing the dictionary.

\begin{figure}[H]
\begin{center}
\includegraphics[width=\textwidth]{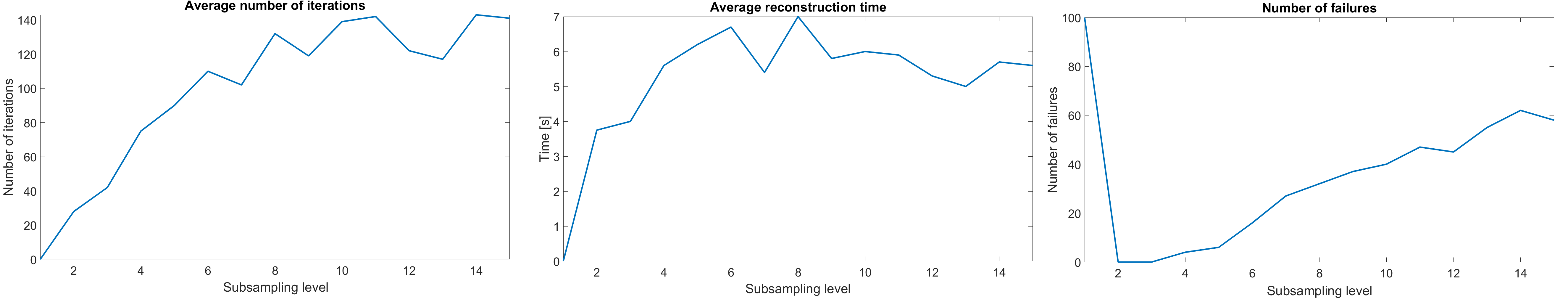}
\end{center}
\caption[width=\textwidth]
{ \label{fig:result_diagrams_sp} Here we see results for different potential sound source position grid subsampling (from no subsampling up to subsampling 15 times).}
\end{figure}

In Figure \ref{fig:result_diagrams_sp} we can see results for no subsampling over resonant frequencies (first 63 resonant frequencies were taken into account - room modes between $(1,0,0)$ and $(3,3,3)$) and different subsamplings over the potential sound source positions. There were no successful reconstruction attempts when the whole grid was taken into account. Subsampling 2 or 3 times showed the best performance with the convergence within the predefined 300 iterations. Average number of iterations and average reconstruction time were computed only for the successful quick reconstructions.

\begin{figure}[H]
\begin{center}
\includegraphics[width=\textwidth]{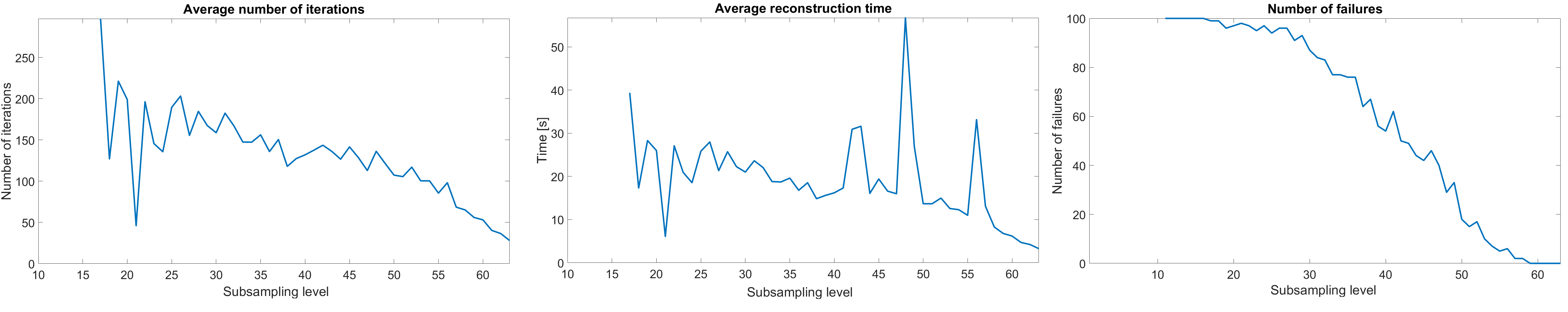}
\end{center}
\caption[width=\textwidth]
{ \label{fig:result_diagrams_rf} Here we see results for different resonant frequency grid subsampling (from selecting 11 up to selecting all 63 resonant frequencies).}
\end{figure}

In Figure \ref{fig:result_diagrams_rf} we can see results for subsampling level of 2 over the potential positions of sound sources and different subsets of resonant frequencies have been taken into account (from 11 up to 63 out of 63). If we choose a subset of below 17 resonant frequencies, the algorithm never converges. If we had average results over more than 100 simulations, the curves in the results would have been smoother. We see that we can not subsample a lot such a small set of resonant frequencies.

Therefore, we have to subsample the sound source position grid since we are dealing with a highly coherent dictionary. By increasing the level of subsampling over columns of the dictionary, we decrease the probability that the atoms that we are searching for are present in the subset. On the other hand, the resonant frequency grid should not be too oversampled in order to achieve a quick convergence (below predefined 300 iterations or similar).

\subsection{Precision and Basis Mismatch}
Due to the smoothness of the room mode functions, there is a small variation in the value between the close points. This supports the idea of similarity of the atoms of the dictionary of the spatially close positions.

Compressed sensing usually assumes the existence of a grid with finite density and our signals of interests can fail to coincide with the nodes of the predefined grid, especially in the case of moving sources. As shown in \cite{BasisMismatch} this can cause that sparse signals appear as incompressible. The work we have observed before \cite{CompressiveBeamforming} has an extension to a continuous case \cite{GridFreeBeamfroming} by applying the semi-definite programming \cite{SDP}. In our observations we have assumed that our grid of the potential positions of the sources is dense enough to avoid the spectral leakage and continuous approaches will be left for future work.

\subsection{Requirements and Limitations}
\par In a setting where we have multiple sound sources and a microphone, the sound received is equal to the linear combination of the convolution of sounds emitted by the sound sources and the transfer functions that correspond to their positions. Therefore we need the following assumption: we can efficiently localize sources which have a flat spectrum in the observed frequency band, since they result in a nearly constant Fourier coefficients of emitted signals spectrum. Otherwise we have to know upfront the signals that will be emitted by sources.

In order to avoid ill-conditioness the microphone should lie off the planes of symmetry.

\section{Conclusion} \label{Conclusion}
By observing the sound source localization problem through the theory of compressed sensing, we have enabled localization of multiple sound sources in a room using only one microphone. Unlike most of the localization algorithms, this approach guaranties the localization in 3D, without neglecting the elevation angle, which is rarely estimated. The simplicity of our solution lays in the low required prior knowledge about the room - only the height of the peaks in the RTF at the resonant frequencies should be know.

Our solution has the potential of being applied to the optimization of the quality of the hearing aids - once the location of source is estimated we can introduce weighting on the reception side, as well as in robotics for monoaural localization. The emerging field of virtual reality would be just another domain of potential application.

Future work will include estimation \textit{off the grid} in order to avoid the basis mismatch and the challenging computational costs. Removal of the assumption on the level of sparsity should also be investigated further.

\section{Supplementary Materials}
\textit{matlab} code used for generating each of the figures in this paper as well as the acoustical room mode framework is available for download on the following link: \url{https://github.com/epfl-lts2/room_transfer_function_toolkit}. \textit{python} version of the toolkit is also available.

\acknowledgments 
The work of H. Peić Tukuljac was supported  by the Swiss National Science Foundation under Grant No. 200021\texttt{\char`_}169360 for the project \textit{“Compressive Sensing applied to the Characterization and the Control of Room Acoustics”}. We would like to thank Adrien Besson for fruitful discussions and valuable suggestions during the preparation of this manuscript.
\bibliography{report} 

\begin{thebibliography}{10}

\bibitem{DonohoCS}
Donoho, D.~L., ``Compressed sensing,'' {\em {IEEE} Trans. Information
  Theory}~{\bf 52}(4),  1289--1306 (2006).

\bibitem{CandesCS}
Candes, E.~J., Romberg, J., and Tao, T., ``Robust uncertainty principles: Exact
  signal reconstruction from highly incomplete frequency information,'' {\em
  IEEE Trans. Inf. Theor.}~{\bf 52},  489--509 (Feb. 2006).

\bibitem{CSAPP}
Boche, H., Calderbank, R., Kutyniok, G., and Vybral, J.,  [{\em Compressed
  Sensing and Its Applications: MATHEON Workshop
  2013}{\nolinebreak\hspace{0.1em}]}, Birkhäuser Basel, 1st~ed. (2015).

\bibitem{CompressiveBeamforming}
Xenaki, A., Gerstoft, P., and Mosegaard, K., ``Compressive beamforming,'' {\em
  The Journal of the Acoustical Society of America}~{\bf 136}(1),  260--271
  (2014).

\bibitem{HearingBehindWalls}
Kitić, S., Bertin, N., and Gribonval, R., ``Hearing behind walls: Localizing
  sources in the room next door with cosparsity,'' in [{\em 2014 IEEE
  International Conference on Acoustics, Speech and Signal Processing
  (ICASSP)}{\nolinebreak\hspace{0.1em}]},   3087--3091 (May 2014).

\bibitem{RoomShape}
Dokmani{\'c}, I., Parhizkar, R., Walther, A., Lu, Y.~M., and Vetterli, M.,
  ``Acoustic echoes reveal room shape,'' {\em Proceedings of the National
  Academy of Sciences}~{\bf 110}(30),  12186--12191 (2013).

\bibitem{EchoLocalization}
X.~Falourd, L.~Rohr, M.~R. and Lissek, H., ``Spatial echogram analysis of a
  small auditorium with observations on the dispersion of early reflections,''
  {\em Inter-Noise 2010 - noise and sustainability}  (June 2010).

\bibitem{WaveRepresentation}
Koyano, Y., Yatabe, K., Ikeda, Y., and Oikawa, Y., ``Physical-model based
  efficient data representation for many-channel microphone array,'' in [{\em
  2016 IEEE International Conference on Acoustics, Speech and Signal Processing
  (ICASSP)}{\nolinebreak\hspace{0.1em}]},   370--374 (March 2016).

\bibitem{Low}
Mignot, R., Chardon, G., and Daudet, L., ``Low frequency interpolation of room
  impulse responses using compressed sensing,'' {\em IEEE/ACM Trans. Audio,
  Speech and Lang. Proc.}~{\bf 22},  205--216 (Jan. 2014).

\bibitem{VehicleLocalization}
Marmaroli, P., Carmona, M., Odobez, J.~M., Falourd, X., and Lissek, H.,
  ``Observation of vehicle axles through pass-by noise: A strategy of
  microphone array design,'' {\em IEEE Transactions on Intelligent
  Transportation Systems}~{\bf 14},  1654--1664 (Dec 2013).

\bibitem{BimodalLocalization}
Marmaroli, P., Odobez, J.~M., Falourd, X., and Lissek, H., ``A bimodal sound
  source model for vehicle tracking in traffic monitoring,'' in [{\em 2011 19th
  European Signal Processing Conference}{\nolinebreak\hspace{0.1em}]},
  1327--1331 (Aug 2011).

\bibitem{RoomAcoustics}
Kuttruff, H. and Mommertz, E.,  [{\em Room
  Acoustics}{\nolinebreak\hspace{0.1em}]},  239--267, Springer Berlin
  Heidelberg, Berlin, Heidelberg (2013).

\bibitem{CurveFitting}
Richardson, M.~H. and Formenti, D.~L., ``Global curve fitting of frequency
  response measurements using the rational fraction polynomial method,''
  (1985).

\bibitem{SparseCompare}
Tropp, J.~A. and Wright, S.~J., ``Computational methods for sparse solution of
  linear inverse problems,'' {\em Proceedings of the IEEE}~{\bf 98},  948--958
  (June 2010).

\bibitem{ConvexRelaxation}
Tropp, J.~A., ``Just relax: convex programming methods for identifying sparse
  signals in noise,'' {\em IEEE Transactions on Information Theory}~{\bf 52},
  1030--1051 (March 2006).

\bibitem{Boyd}
Boyd, S. and Vandenberghe, L.,  [{\em Convex
  Optimization}{\nolinebreak\hspace{0.1em}]}, Cambridge University Press, New
  York, NY, USA (2004).

\bibitem{l1morel0}
Cand{\`e}s, E.~J., Wakin, M.~B., and Boyd, S.~P., ``Enhancing sparsity by
  reweighted l 1 minimization,'' {\em Journal of Fourier Analysis and
  Applications}~{\bf 14}(5),  877--905 (2008).

\bibitem{OMP}
Tropp, J.~A., Gilbert, A.~C., and Strauss, M.~J., ``Algorithms for simultaneous
  sparse approximation: Part i: Greedy pursuit,'' {\em Signal Process.}~{\bf
  86},  572--588 (Mar. 2006).

\bibitem{CoSaMP}
Needell, D. and Tropp, J., ``Cosamp: Iterative signal recovery from incomplete
  and inaccurate samples,'' {\em Applied and Computational Harmonic
  Analysis}~{\bf 26}(3),  301 -- 321 (2009).

\bibitem{FRI}
Blu, T., Dragotti, P.~L., Vetterli, M., Marziliano, P., and Coulot, L.,
  ``Sparse sampling of signal innovations,'' {\em IEEE Signal Processing
  Magazine}~{\bf 25},  31--40 (March 2008).

\bibitem{BasisMismatch}
Chi, Y., Pezeshki, A., Scharf, L., and Calderbank, R., ``Sensitivity to basis
  mismatch in compressed sensing,'' in [{\em 2010 IEEE International Conference
  on Acoustics, Speech and Signal Processing}{\nolinebreak\hspace{0.1em}]},
  3930--3933 (March 2010).

\bibitem{CoherentDictionaryRecovery}
Candès, E.~J., Eldar, Y.~C., and Needell, D., ``Compressed sensing with
  coherent and redundant dictionaries,'' {\em CoRR}~{\bf abs/1005.2613} (2010).

\bibitem{RedundantDictionaries}
Schnass, K. and Vandergheynst, P., ``Dictionary preconditioning for greedy
  algorithms,'' {\em IEEE Transactions on Signal Processing}~{\bf 56},
  1994--2002 (May 2008).

\bibitem{RandomSubdictionaries}
Tropp, J.~A., ``On the conditioning of random subdictionaries,'' {\em Applied
  and Computational Harmonic Analysis}~{\bf 25}(1),  1 -- 24 (2008).

\bibitem{MPSto}
Peel, T., Emiya, V., Ralaivola, L., and Anthoine, S., ``Matching pursuit with
  stochastic selection,'' in [{\em 2012 Proceedings of the 20th European Signal
  Processing Conference (EUSIPCO)}{\nolinebreak\hspace{0.1em}]},   879--883
  (Aug 2012).

\bibitem{StoCoSaMP}
Pal, D.~K. and Mengshoel, O.~J., ``Stochastic cosamp: Randomizing greedy
  pursuit for sparse signal recovery,'' in [{\em Machine Learning and Knowledge
  Discovery in Databases - European Conference, {ECML} {PKDD} 2016, Riva del
  Garda, Italy, September 19-23, 2016, Proceedings, Part
  {I}}{\nolinebreak\hspace{0.1em}]},   761--776 (2016).

\bibitem{GridFreeBeamfroming}
Xenaki, A. and Gerstoft, P., ``Grid-free compressive beamforming,'' {\em
  CoRR}~{\bf abs/1504.01662} (2015).

\bibitem{SDP}
Vandenberghe, L. and Boyd, S., ``Semidefinite programming,'' {\em SIAM
  Rev.}~{\bf 38},  49--95 (Mar. 1996).

\end{thebibliography}
\bibliographystyle{spiebib} 

\end{document}